\documentstyle[preprint,prd,aps,psfig,floats]{revtex}
\newcommand{\be}{\begin{equation}}
\newcommand{\ee}{\end{equation}}
\newcommand{\bea}{\begin{eqnarray}}
\newcommand{\eea}{\end{eqnarray}}
\renewcommand{\d}{{\rm d}}

\tightenlines
\begin{document}
\draft
%
\firstfigfalse
%
\title{Equilibration of the Gluon-Minijet Plasma at RHIC and LHC}
\author{Gouranga C. Nayak$^a$, Adrian Dumitru$^b$, Larry McLerran$^c$,
Walter Greiner$^a$}
\address
{$^a$\small\it{Institut f\"ur Theoretische Physik,
J. W. Goethe-Universit\"at,
60054 Frankfurt am Main, Germany}}
\address{$^b$Department of Physics, Columbia University,
        538W 120th Street, New York, NY 10027, USA
        }
\address
{$^c${\small\it Physics Department,  Brookhaven National Laboratory
        Upton, NY 11979, USA  }}     

\maketitle

\begin{abstract} 
We study the production and equilibration of the gluon-minijet plasma expected 
to be formed in the central region of ultrarelativistic heavy-ion
collisions at the BNL-RHIC and the CERN-LHC by solving a self-consistent
relativistic transport equation.
We compute the minijet production within perturbative QCD. 
Subsequent collisions among the
semi-hard partons are treated by considering the elastic 
$gg \rightarrow gg$ processes with screening of the long wavelength 
modes taken into account. We determine the time $\tau_{eq}$
where close to ideal
hydrodynamic flow sets in, and find rather similar numbers for
central heavy-ion collisions at BNL-RHIC and CERN-LHC energies.
The number densities, 
energy densities and temperatures of the minijet plasma
are found to be different at RHIC and LHC, e.g.\
$T(\tau_{eq})\sim220$~MeV and 380~MeV, respectively.
\end{abstract}  
\bigskip

%
\section{Introduction}
In the next few years, the BNL-RHIC (Au-Au collisions at $\sqrt{s}$=200 GeV per
incident nucleon pair) and the CERN-LHC (Pb-Pb collisions at
$\sqrt{s}$=5.5A TeV)
accelerators will provide the opportunity to study a new phase of 
matter, namely the so-called Quark-Gluon Plasma (QGP)~\cite{QM99}. 
It is very important and interesting to study whether the QGP
actually does thermalize in those reactions, and if so, what is the 
actual energy density, number density and temperature at which it thermalized.
For this purpose, and also for the calculation of all the signatures,
it is necessary to study the space-time evolution of partons just after the
nuclear collision. For example, the equilibration time is crucial for a
quantitative understanding of $J/{\psi}$ suppression~\cite{nayak},
and it is a challenging task to 
determine this quantity accurately. Similarly, understanding the equilibration
time is important for all other proposed signatures for the QGP, for example
for dilepton emission and strangeness production~\cite{QM99}. Once equilibrium 
is reached, the further space time evolution of partons can be described 
by the well known equations of hydrodynamics.

The evolution of the QGP towards (local) equilibrium can be studied 
by solving transport equations for quarks and gluons with all the dynamical
effects taken into account. Obviously, the first problem one always encounters 
is the correct computation of the initial conditions needed to
solve the transport equation. This is because one can not calculate the
parton production in all range of momentum from perturbative QCD (pQCD).
There are also 
coherence effects~\cite{mclerran,mueller} that play an important role 
in the early stage of the nuclear collision at very high energy. For small 
x and large nuclei, the QCD based calculation performed in~\cite{mclerran1}
predicts
the existence of a coherent field in a certain kinematical range.
That field may play an important role in the equilibration of the plasma.
In the present paper, however, we restrict our calculation to the initial 
incoherent parton production, which is computed within the framework of pQCD.
We study the subsequent evolution of that minijet ``plasma''  by solving a
relativistic transport equation, thus taking into account collisions between
the produced partons. In the future, we intend to generalize our approach
to include both coherent field and incoherent
partons in the transport equation.

The paper is organized as follows. In section II we briefly review
minijet production in high-energy nuclear collisions within pQCD.
In section III we discuss the in-medium screening of long wavelength gluons
which is relevant to our study. We present the relativistic 
transport equation in section IV, briefly
describing the numerical strategy for its solution in section V. 
We discuss our main results in section VI and conclude in section VII.
Throughout the manuscript we employ natural units, $\hbar=c=k=1$.

\section{Minijet Production in nuclear collisions at RHIC and LHC}
\label{mj_prod}
In this section we review the computation of the single-inclusive
semi-hard cross section in lowest order pQCD, cf.\ also~\cite{edens,minij}.
The $2 \rightarrow 2$ minijet cross
section per nucleon in AA collision is given by 
\begin{equation}
\sigma_{jet} = \int dp_t dy_1 dy_2 {{2 \pi p_t} \over {\hat{s}}} 
\sum_{ijkl}
x_1~ f_{i/A}(x_1, p_t^2)~ x_2~ f_{j/A}(x_2, p_t^2)~
\hat{\sigma}_{ij \rightarrow kl}(\hat{s}, \hat{t}, \hat{u}).
\label{jet}
\end{equation}
Here $x_1$ and $x_2$ are the light-cone momentum fractions carried by
the partons $i$ and $j$ from the projectile and the target,
respectively. $f_{j/A}$ are the distribution functions of the parton
species $j$ within a nucleon bound in a nucleus of mass number $A$.
$y_1$
and $y_2$ denote the rapidities of the scattered partons. The symbols
with carets refer to the parton-parton c.m.\ system.
$\hat{\sigma}_{ij \rightarrow kl}$ is the elementary pQCD parton cross
section. 
\begin{equation}
\hat{s}=x_1 x_2 s = 4 p_t^2 ~{\rm cosh}^2
\left ( {{y_1-y_2} \over {2}} \right ),
\end{equation}
gives the total c.m.-energy of the parton-parton scattering.
The rapidities $y_1$, $y_2$ and the momentum fractions $x_1$, $x_2$ are
related by,
\begin{equation}
x_1=p_t~(e^{y_1}+e^{y_2})/{\sqrt{s}}, \hspace{0.5cm}
x_2=p_t~(e^{-y_1}+e^{-y_2})/{\sqrt{s}}.
\end{equation}
The limits of integrations of rapidities $y_1$ and $y_2$ are given by
$\vert {y_1} \vert \leq$ ln($\sqrt{s}/{2 p_t} + \sqrt{s/{4 p_t^2}
-1}$) and $-$ln(${\sqrt{s}/{p_t}-e^{-y_1}}) \leq y_2 \leq$
ln(${\sqrt{s}/{p_t}-e^{y_1}})$, respectively.
We multiply the above minijet cross
sections by the phenomenological factor $K=2$ to account for higher-order
contributions.

The minijet cross section, Eq.\ (\ref{jet}),
can be related to the total number of produced partons via
\begin{equation}
N=T(0) ~\int dp_t dy_1 dy_2 {{2 \pi p_t} \over {\hat{s}}} 
\sum_{ijkl}
x_1~ f_{i/A}(x_1, p_t^2)~ x_2~ f_{j/A}(x_2, p_t^2)~
\hat{\sigma}_{ij \rightarrow kl}(\hat{s}, \hat{t}, \hat{u}),
\label{number}
\end{equation}
where $T(0)= 9A^2/{8\pi R_A^2}$ is the nuclear geometrical factor for
head-on AA collisions (for a nucleus with a sharp surface).
$R_A=1.1 A^{1/3}$~fm is the nuclear radius.
Similarly, the total transverse energy $E$ of minijets is given by
\begin{equation}
E=T(0) ~\int dp_t p_t dy_1 dy_2 {{2 \pi p_t} \over {\hat{s}}} 
\sum_{ijkl}
x_1~ f_{i/A}(x_1, p_t^2)~ x_2~ f_{j/A}(x_2, p_t^2)~
\hat{\sigma}_{ij \rightarrow kl}(\hat{s}, \hat{t}, \hat{u}).
\label{energy}
\end{equation}
We employ the ``EKS98'' set of parton distribution functions
in a bound nucleon from ref.~\cite{eskola1}, based on
the GRV98 set of parton distributions for a free
nucleon~\cite{grv98}.
One has to choose the scale in the transverse momentum below which the 
incoherent parton picture is not valid and coherence effects have to 
be taken into account. These scales provide the initial cutoff for
the (semi-)hard scatterings. The values for the cutoff are found to be 
$\sim$ 1 at RHIC and $\sim$ 2 GeV at LHC from the McLerran-Venugopalan 
model using their initial conditions \cite{mueller}. 
For the quantitative estimates presented below
we actually choose the cutoff
$p_0$ to be 1.13~GeV and 2.13~GeV at RHIC and LHC, respectively. 
These values are obtained by an independent method by Eskola 
{\it et. al,} \cite{cutesk}.

To solve the transport equation one has to specify the initial distribution
function of the particles in phase-space,
while accounting for the correlation between 
momentum-space rapidity and space-time rapidity. We choose our 
initial distribution function to be a Boltzmann distribution in the
local rest frame,
\be
f(\tau_0=1/p_0, \xi,p_t)=\exp\left(-p_\mu u^\mu/T_{jet}\right) =
\exp\left(-p_t \cosh{\xi}/T_{jet}\right).
\label{f0}
\ee
$u^\mu=(\cosh\eta,0,0,\sinh\eta)$ is the four-velocity of the local rest-frame
of the medium, with $\eta={\rm Artanh}\,(z/t)$ denoting
the space-time rapidity, and $\xi\equiv\eta-y$. 
The parameter $T_{jet}$ can be determined from the average energy per 
particle of the initially formed minijets which is obtained from 
Eqs.~(\ref{energy}) and (\ref{number}). We mention here that
even though the initial distribution is chosen to be a Boltzmann distribution,
the subsequent evolution will be non-ideal, see below. 

We shall use these initial 
conditions to study the evolution of the parton ``plasma'' at RHIC and
LHC by solving a self-consistent relativistic transport equation
(see section IV). As most of the produced minijets are gluons, 
we will simplify our considerations by considering gluons only.
In Table-I we have listed the initial conditions of the gluon
minijets.

\section{Screening in Non-Equilibrium}
In this section we describe the screening of long wavelength electric fields
in the parton medium. It will play an important role in defining the finite
collision term of our transport equation. We shall employ the
static limit (screening of infinitely long wavelength fields), which gives
the Debye screening mass, and assume that this simplified treatment is at
least qualitatively correct. Our concern here is to incorporate the
density dependence of the medium-induced cutoff (which thus also depends on
collision energy), as well as to treat the coupled evolution (cutoff and
medium) self-consistently. At present we can not include magnetic
(dynamic) screening into the non-equilibrium study because 
we do not find an expression for the magnetic screening mass in terms
of the non-equilibrium distribution function in the literature (see section VII
for a detail discussion). We also mention here that the
momentum dependent screening mass is usually studied in thermal equilibrium.
However, accounting for the momentum dependence of the screening mass in 
non-equilibrium is technically very difficult and is beyond
the scope of this present paper. 

The electric screening mass is given by the infrared limit of the real part 
of the gluon self-energy $\Pi^{00}$,
calculated in the given background that is 
described by the distribution function $f$ (not to be confused with the
parton distribution function $f_{j/A}$ introduced in section II).
In ref.~\cite{biro} the following expression has been derived (in Coulomb
gauge) for a medium of gluonic excitations:
\be 
m^2= - \frac{3\alpha_s}{\pi^2} \lim_{|\vec{q}|\rightarrow0}
\int d^3p\, \frac{|\vec{p}|}{\vec{q}\cdot\vec{p}}\, \vec{q}\cdot\nabla_p
f(p).
\label{mscr}
\ee
In the above equation $\vec{q}$ is the momentum of the test particle, 
$f(p)$ is the non-equilibrium distribution function of the gluons and
$\alpha_s$ is the strong coupling constant. 
We will consider the transverse screening mass in the following, which will
be introduced below as a cutoff in parton-parton anti-collinear
elastic scattering to
obtain a finite transport cross-section.
Performing an integration by parts
we obtain the transverse screening mass
\be 
m_t^2 = \frac{3\alpha_s}{\pi^2} \int \frac{\d^3 p}{p^0} f(p).
\label{mscr1}
\ee
Following Bjorken's hypothesis~\cite{bjorken}, we express all quantities
in terms of longitudinal-boost invariant parameters $\tau=\sqrt{t^2-z^2}$,
$\xi$ and $p_t$. We assume
that the above equation is also valid for a space-time dependent distribution
function $f(x,p)$ and hence use
\be \label{m_t_scr}
m_t^2 (\tau)=
\frac{6\alpha_s(\tau)}{\pi} \int dp_t p_t\int d\xi \, f(\tau,p_t,\xi)
 \quad.
\label{mt}
\ee
Improving earlier approaches~\cite{Biro:1993qt}, we do not assume
factorization of the distribution function in the form
$f(\tau,p_t,\xi)=g(\xi)h(\tau,p_t)$. Also, in our case the screening
mass enters the collision kernel of the transport equation and thus
determines the rate of equilibration.

In the above equation the QCD 
running coupling constant becomes time dependent, via
\be
\alpha_s(\tau)=\alpha_s(\langle p_t^2(\tau)\rangle).
\label{alfa}
\ee
The average transverse momentum squared of the excitations of the medium
is defined as
\be \label{av_pt2}
\langle p_t^2(\tau)\rangle
 = \frac{\int d\Gamma p_{\mu}u^{\mu} p_t^2 f(\tau,p_t,\xi)}
 {\int d\Gamma p_{\mu}u^{\mu} f(\tau,p_t,\xi)}.
\label{pt2}
\ee
Here $\d\Gamma={\d^3p}/{{(2\pi)}^3p^0} ={\d p_t p_t \d\xi}/{{(2\pi)}^2}$ 
is the invariant momentum-space measure. Throughout the manuscript we use 
the GRV98 calculation of $\alpha_s(\langle p_t^2(\tau)\rangle)$ \cite{grv98} 
with $\langle p_t^2(\tau)\rangle$ calculated from Eq. (\ref{pt2}).

\section{Solution of the Transport Equation with Screening}

In the absence of any coherent color field the space-time evolution of the 
produced partons at RHIC and LHC can be studied by solving the Boltzmann
transport equation
\be
p^{\mu} \partial_{\mu}f(x,p)=C(x,p),
\ee
where $f(x,p)$ is the distribution function of gluons and $C(x,p)$ is the
collision term. To solve the above transport equation with the initial value
of $f_0$ given by Eq. (\ref{f0}), we employ the relaxation time approximation
for the collision term~\cite{Baym,HosKaj}:
\be \label{t-eqn}
C(\tau,\xi,p_t) = -p^\mu u_\mu \left[
f(\tau,\xi,p_t)-f^{eq}(\tau,\xi,p_t)\right] / {\tau_c(\tau)}.
\ee
$f^{eq}$ is the Bose-Einstein equilibrium
distribution function, and $\tau_c(\tau)$ is the time dependent 
relaxation time of the plasma. With this collision term the formal 
solution of the transport equation becomes
\bea
f(\tau,\xi,p_t) = \int_{\tau_0}^{\tau} \d \tau^\prime 
\exp \left(\int_{\tau}^{\tau^{\prime}} 
\frac{\d\tau^{\prime\prime}}{\tau_c(\tau^{\prime\prime})}
 \right) \frac{f^{eq}(\tau^\prime,\xi^\prime,p_t)}
{\tau_c(\tau^{\prime})} +~f(\tau_0, \xi_0, p_t)~
\exp\left({-\int_{\tau_0}^{\tau} \frac{\d\tau^{\prime\prime}}
{\tau_c(\tau^{\prime\prime})}}\right),
\label{dist}
\eea
where $\xi^{\prime}$ is the solution of
\begin{equation}
\sinh\xi^\prime=  \frac{\tau}{\tau^\prime} \sinh \xi~,
\end{equation}
and $\sinh\xi_0=\frac{\tau}{\tau_0} \sinh \xi$, with $\tau_0=1/p_0$. 
Despite the fact that the transport cross section (governing kinetic
equilibration) has been derived from the $gg\rightarrow gg$ elastic scattering
cross section, the number-current of the gluons is not conserved.
Rather, for massless particles the number density in equilibrium
is proportional to the entropy density. However, dissipation during the
pre-equilibrium
stage will produce additional entropy, and accordingly the comoving number
density is expected to decrease less fast than for a conserved current 
$j^\mu=\rho u^\mu$, where
$\partial_\mu(\rho u^\mu)=\d\rho/\d\tau+\rho/\tau=0$, see below.

We write the relaxation time for collisions, $\tau_c(\tau)$, as
\be \label{tauc_sigmat}
\tau_c(\tau) = \frac{1}{\sigma_t(\tau) n(\tau)} \quad,
\label{tauc}
\ee
where 
\be
n(\tau) = g_G \int\d \Gamma p_{\mu}u^{\mu} f(\tau,p_t,\xi)
\label{den}
\ee
is the number density of the gluon-minijet plasma, and
\be 
\sigma_t(\tau) = \int \d \Omega \frac{\d\sigma}{\d \Omega}
\sin^2 \theta_{c.m.}
\label{trXsection}
\ee
denotes the time dependent transport cross-section for the collision 
processes \cite{HosKaj,Danielewicz:1984ww}. We assume that anti-collinear
small-angle scattering gives the dominant contribution to the transport
cross-section~\cite{Danielewicz:1984ww}, such that
$\sin^2\theta_{c.m.}=4\hat{t}\hat{u}/\hat{s}^2$.
We mention again that, in our study, all the quantities such as
$m_t$, $\sigma_t$, $n$, $\tau_c$ are time dependent and have been obtained
from the non-equilibrium distribution function of the gluon minijets.

We shall consider the leading order elastic scattering processes 
$gg \rightarrow gg$. The differential cross section for this
process is given by
\be
\frac{\d\sigma}{\d\hat{t}} = \frac{9\pi \alpha_s^2}{2{\hat{s}}^2} 
\left[3-\frac{\hat{u}\hat{t}}{{\hat{s}}^2}-\frac{\hat{u}\hat{s}}{{\hat{t}}^2}
-\frac{\hat{s}\hat{t}}{{\hat{u}}^2}\right].
\ee
In the limit of small-angle scattering (of identical particles)
it simplifies to
\be
\frac{\d\sigma}{\d\hat{t}} = \frac{9\pi \alpha_s^2}{2}\frac{1}{{\hat{t}}^2}
\ee
and the transport cross-section
$\sigma_t$ diverges logarithmically due to exchange of long-wavelength
gluons. However, as discussed in section III,
long-wavelength fields will be screened by the dense
medium. For our studies we therefore employ the
medium-modified elastic cross-section~\cite{Gyulassy:1997ib}
\be
\frac{\d \sigma}{\d \hat{t}} = \frac{9\pi\alpha_s^2}{2} 
\left(\frac{m_t^2}{\hat{s}}+1 \right)
\frac{1}{\left(\hat{t}-m_t^2\right)^2} \quad.
\label{mdscr}
\ee
Using Eq.~(\ref{mdscr}) in Eq.~(\ref{trXsection}) we obtain the medium 
modified finite (time dependent) transport cross-section 
\bea
\sigma_t(\tau) = \frac{9}{2} \frac{4\pi\alpha_s^2(\tau)}{\hat{s}^2(\tau)}
\left(\frac{m_t^2(\tau)}{\hat{s}(\tau)}+1 \right)
\left[ \left(\hat{s}(\tau)+2m_t^2(\tau)\right)\log\left(\frac{\hat{s}(\tau)}
{m_t^2(\tau)}+1 \right)
-2\hat{s}(\tau)\right] \quad.
\label{cros}
\eea
To simplify the considerations we have replaced
$\hat{s}$ by its average value
\be
\hat{s}(\tau) = 4 \langle E(\tau)\rangle^2
\label{ms}
\ee
in the above expression.
$\langle E(\tau)\rangle$ is the time dependent average energy per 
particle given by 
\be
\langle E(\tau)\rangle = \frac{\epsilon(\tau)}{n(\tau)},
\ee
where
\be
\epsilon(\tau) = g_G \int\d \Gamma (p_{\mu}u^{\mu})^2 f(\tau,p_t,\xi)
\label{end}
\ee
is the energy density of the minijet ``plasma'' and $n(\tau)$ is the local 
number density as defined in Eq.~(\ref{den}).

To obtain the collision term~(\ref{t-eqn}) in each time step we also have to
determine the equilibrium distribution towards which the evolution is
supposed to converge. In other words, we have to determine the ``equivalent''
plasma temperature from~(\ref{end}). It can be obtained from the condition
that the first moment of the actual distribution function $f$ equals that
of the equilibrium distribution function $f^{eq}$, i.e.\
\be
\epsilon = g_G \frac{\pi^2}{30}T^4.
\label{temp}
\ee
This should be a reasonable approximation as long as the system is not
very close to the hadronization phase transition.

\section{Numerical Solution}
The expression for the distribution function $f(\tau, \xi, p_t)$,
Eq.~(\ref{dist}), involves $T(\tau)$ and $\tau_c(\tau)$ which are 
again defined through the distribution function $f(\tau,\xi,p_t)$. 
We solve these coupled set of equations self-consistently.
At any time $\tau$ we start with the old trial values 
$T_O$, $n_O$, ${\alpha_s}_O$, ${m_t}_O$ and ${\hat{s}}_O$ from which we
get $f_O$ via Eq.~(\ref{dist}). This $f_O$ is used in Eq.~(\ref{end})
to calculate $\epsilon(\tau)$ which gives a new temperature $T_N$ through
Eq.~(\ref{temp}). This new temperature, but the old values of 
$n_O$, ${\alpha_s}_O$, ${m_t}_O$ and ${\hat{s}}_O$ are again used in
Eq.~(\ref{dist}) to get a new $f_1$. This $f_1$ is used in Eq.~(\ref{den}) to
calculate a new $n_N$ which also gives a new value ${\hat{s}}_N$ via
Eq.~(\ref{ms}). These new values $T_N$, $n_N$ and ${\hat{s}}_N$ and the 
old values ${\alpha_s}_O$ and ${m_t}_O$ are again used in Eq.~(\ref{dist}) 
to get a new $f_2$. Using this $f_2$ in Eq.~(\ref{pt2}) we obtain
${\alpha_s}_N$ from Eq.~(\ref{alfa}). These new values $T_N$, $n_N$, 
${\hat{s}}_N$, ${\alpha_s}_N$ and old value ${m_t}_O$ are again used in 
Eq. (\ref{dist}) to obtain a new $f_3$. Using this $f_3$ in Eq.~(\ref{mt})
we obtain a new ${m_t}_N$. 
Thus, starting with the old set of values 
$T_O$, $n_O$, ${\alpha_s}_O$, ${m_t}_O$ and ${\hat{s}}_O$
we obtain a new set of values
$T_N$, $n_N$, ${\alpha_s}_N$, ${m_t}_N$ and ${\hat{s}}_N$.
This process is iterated until convergence is attained to the 
required accuracy. This gives us the self-consistent values of 
$\epsilon(\tau)$, $T(\tau)$, $n(\tau)$, ${\alpha_s}(\tau)$, 
${m_t}(\tau)$ and ${\hat{s}}(\tau)$ at any time $\tau$.

\section{Results and Discussions}

The purpose of this paper is to study various bulk properties of the gluon
minijet plasma. In particular,
we discuss the time evolution of
the number density, the energy density, the temperature, as well as the
collision-relaxation time and the transport cross section.
Calculations of signatures 
from this equilibrating minijet plasma will be presented elsewhere. 

We compare the time evolution of the energy density with
that obtained in the free streaming limit (no collisions and initial
rapidity distribution $\d N/\d y\propto\delta(y-\eta)$) and
in the equilibrium limit (isotropic momentum distribution in the comoving
frame, at all times), i.e.\ the hydrodynamical
evolution. For purely longitudinal
expansion the energy density in the free streaming limit
behaves as
\be
\epsilon(\tau) = \epsilon_i{\left(\frac{\tau}{\tau_{fs}}\right)}^{-1}
      \label{efr}
\ee
where $\epsilon_i$ is the energy density of the gluon-minijet plasma
at the time $\tau_{fs}$ where free streaming sets in.

In the equilibrium limit we have
\bea
n(\tau) &=& n_0{\left(\frac{\tau}{\tau_{eq}}\right)}^{-1}, \label{neq}\\
\epsilon(\tau) &=& \epsilon_0{\left(\frac{\tau}{\tau_{eq}}\right)}^{-\frac{4}{3
}},
      \label{eeq}\\
T(\tau) &=& T_0{\left(\frac{\tau}{\tau_{eq}}\right)}^{-\frac{1}{3}}. 
\label{teq}
\eea
In the above equations  $\epsilon_0$, $n_0$ and $T_0$ are the 
energy density, number density and temperature of the gluon plasma 
at $\tau=\tau_{eq}$, where equilibrium is reached. The latter two
equations actually depend on the equation of state of the minijet plasma;
we have assumed an ideal gas of gluons.

The evolution of the energy densities at RHIC and LHC are shown in 
Fig.~1 and Fig.~2 respectively. 
The solid lines depict the result from our self-consistent transport
calculations. The dashed lines
correspond to the free streaming evolution (see Eq. (\ref{efr})) 
with the minijet initial conditions at $\tau_{fs}=\tau_0=1/p_0$. 
The dot dashed lines correspond to the 1$\bigoplus$1 hydrodynamical evolution
of the energy densities (see Eq. (\ref{eeq})) with the 
same initial conditions of the minijet plasma at $\tau_{eq}=\tau_0$. 
The latter case represents the evolution of the energy densities 
assuming equilibration at $\tau=\tau_0$. It can be seen that our results lie
between the free streaming and hydrodynamic limits. 
To see when equilibration sets in we fit our results
with $\epsilon(\tau) \propto \tau^{-\alpha}$, at different times. 
We present our fitted values of $\alpha$ in Table-II and Table-III
for RHIC and LHC respectively. 
It can be observed that as time progresses the scaling exponent
$\alpha$ increases. This is due to the fact that at later times
the collision rate among the plasma constituents increases,
driving the system towards equilibrium. As a cross-check we also
examine the behavior of the temperatures and number densities.

The time evolutions of the ``temperatures'' of
the minijet plasma at RHIC and LHC are depicted in Fig.~3.
It can be observed that the plasma is found to be much hotter at LHC
than at RHIC. To study equilibration we have fitted our results
with $T(\tau) \propto \tau^{-\alpha}$, at different times. 
The fitted values of $\alpha$ are given in Table-II and Table-III
for RHIC and LHC respectively. 
Again we see that as time progresses the scaling exponent $\alpha$ increases. 
As already mentioned above, $T$ should not be interpreted
as a temperature in the thermodynamic sense for $\tau<\tau_{eq}$.
This is because the energy-momentum tensor at those early times deviates from
the ideal fluid form, which is why Eqs.~(\ref{neq}-\ref{teq}) do not hold.
Nevertheless, $T(\tau)$ can be taken as an indication for the time evolution of
the average energy per particle even at $\tau<\tau_{eq}$.

The evolutions of the number densities are shown in Fig.~4.
Not surprisingly, the plasma is found to be much 
denser at LHC than at RHIC. Again, we have fitted our results to
$n(\tau) \propto \tau^{-\alpha}$, at various times. 
The fitted values of $\alpha$ can be found in Table-II and Table-III
for RHIC and LHC respectively. As already mentioned above, the density
$n(\tau)$ decreases less fast than $\propto1/\tau$.

The scaling exponents given in Table-II and Table-III represent the
fitted values of our results ($\epsilon(\tau)$, T($\tau$), n($\tau$))
with the functional form $\tau^{-\alpha}$, at different times.
For an equilibrated $1\bigoplus1$ dimensionally expanding plasma 
the scaling exponents are 4/3 for $\epsilon(\tau)$, 1/3 for T($\tau$)
and 1 for n($\tau$) respectively. It can be seen from Table-II and Table-III
that our scaling exponents approach those values at later times. 
The reader may judge himself when the system is close to
equilibrium. Our choice is $\tau_{eq} \sim 4-5$~fm.
 
The time evolutions of the transport cross sections
at RHIC and LHC are shown in Fig.~5.
In our calculation the minijet scale $p_0$ evolves with collision energy,
(1.13 and 2.13 GeV at RHIC and LHC, respectively; see section~\ref{mj_prod}).
Therefore, more energetic partons are 
produced at LHC and the transport cross section is much smaller. 
These transport cross sections
play a crucial role in the equilibration of the plasma, cf.\
Eq.~(\ref{tauc}). Since the screening mass and the momentum 
scale in the running coupling constant decrease with time the
transport cross section, in turn, increases strongly. Therefore,
despite the decreasing number density the relaxation time
decreases at later time. In Fig.~6 
we have displayed the time evolution of the relaxation time 
$\tau_c(\tau)$. One observes that the relaxation times 
at RHIC and LHC do not differ by much, despite the much higher density
of partons obtained at LHC. 

Finally, in Fig.~7 we present the time evolution of the coupling constant.
One observes that the coupling seems to be weak at LHC but becomes 
stronger at RHIC, see also ref.~\cite{Bass:1999bu}. One may have to include
also higher order processes in the study of equilibration of the minijet plasma
at RHIC, which will yield faster equilibration as compared to the
present results.

\section{Summary and Conclusions}

We have studied the production and equilibration of the gluon minijet plasma
produced in the central region of high-energy nuclear collisions
by solving the relativistic Boltzmann transport equation. 
The initial conditions are obtained from pQCD. We have solved the
transport equation employing a collision term based on
$2 \rightarrow 2$ elastic collisions.
The collinear divergence in the perturbative cross section to lowest order
is removed by incorporating the screening of very soft interactions
by the medium. The
screening mass is calculated from the non-equilibrium distribution
function of the partons.

Parton equilibration in high-energy collisions has previously been
discussed within the Boltzmann equation in
refs.~\cite{heiselberg1,heiselberg2,wong1}.
However, ref.~\cite{heiselberg1} made an {\sl Ansatz} for the
time-dependence of the relaxation rate,
$\tau_c(\tau) = \theta_0 (\tau/\tau_0)^{\beta}$. It has been found that
the results depend rather sensitively on the exponent $\beta$. 
In~\cite{heiselberg2} an attempt was made to 
obtain an analytical expression for $\tau_c(\tau)$. However, the authors
actually use equilibrium distribution functions (apart
from other approximations) to derive the relaxation
time, which is written as a function of temperature.

In the present
work we attempted to reduce such uncertainties by coupling $\tau_c(\tau)$ to
the evolution of the medium itself, leading to self-consistent
dynamics. Moreover, $\tau_c(\tau)$ is determined from the medium
by using the number density and transport cross section which are
computed by using the actual non-equilibrium distribution functions. 
The treatment of a non-equilibrium medium determining the relaxation time
in an explicit and numerically computable way is one of the main 
contributions of our paper. 

Another major difference to previous treatments
lies in the choice of the initial time and initial conditions.
In particular, we have taken into account that according to recent
arguments regarding parton saturation the minijet scale increases with
collision energy, which has a significant effect on the initial conditions.
In~\cite{heiselberg1,heiselberg2,wong1}, for example, the initial
conditions were taken from the HIJING model~\cite{hijing} 
at rather late time where it is assumed that momentum isotropy is reached 
(in HIJING the minijet scale $p_0$ is assumed to be energy independent). 
The initial conditions used in the above studies are obtained at $\tau'_0$=
0.7~fm at RHIC and 0.5~fm at LHC. The energy densities at these
times are 3.2~GeV/fm$^3$ at RHIC and 40~GeV/fm$^3$ at LHC. We solve the
transport equation starting at $\tau_0=1/p_0<\tau'_0$~fm,
and do not find isotropic momentum distributions at $\tau'_0$.
As our initial conditions
at $\tau_0=1/p_0$ are rather different from those employed
in~\cite{heiselberg1,heiselberg2,wong1},
our findings regarding equilibration time and other bulk properties 
of the plasma differ considerably from those of the above studies.

In our study all the quantities such as $\alpha_s(\tau)$, $m_t(\tau)$,
$\hat s(\tau)$, $n(\tau)$, $\epsilon(\tau)$ are obtained by using the
non-equilibrium distribution functions. These time dependent quantities
are used in a self-consistent manner
to determine the evolution of the plasma. 
We have attempted to study equilibration
without approximations which are valid only in equilibrium, and have tried to
define most of the quantities in non-equilibrium. For example,
we do not include
a magnetic screening mass in our study because we do not find
a closed expression for the magnetic screening mass in terms of the
non-equilibrium distribution function in the literature. Most of the
analysis of magnetic screening mass is done in thermal QCD and such
calculations can not be applied to the initial non-equilibrium
stage of high-energy collisions. From this point of view
we are presently unable to account for magnetic screening in 
our non-equilibrium study. Also, the extensive numerical computations
(we solve for the actual phase-space distribution function, not only
for it's first moment) were presently restricted to a momentum independent
relaxation time; that is, we considered
the static (infinite wavelength) limit of the polarization tensor only.

In the present discussion of thermal equilibration 
we have restricted ourselves to $gg \rightarrow gg$ secondary elastic
collisions. We hope to include $gg \rightarrow ggg$ (and vice versa)
inelastic collisions in the future, which is necessary to address chemical
equilibration. 
Our present results are based on $gg \rightarrow gg$ perturbative
elastic collisons of the produced minijets at RHIC and LHC.

The present study confirms that the plasma is much denser and hotter
at LHC energy. For central collisions of heavy ions at RHIC and LHC 
energy the minijet plasma appears close to equilibrium at a time
$\tau_{eq}\simeq4-5$~fm, with a 
temperature of 220~MeV and 380~MeV respectively. We do not obtain
significantly different kinetic
equilibration times of the gluon-minijet plasma at RHIC and LHC because the
product of comoving parton density and transport cross-section, i.e.\
the kinetic equilibration rate, is similar.
However, there are
differences in other physical quantities such as number density, energy 
density and temperature. Our results indicate
somewhat larger equilibration times than those obtained in the parton cascade
model~\cite{pcm} (with fixed minijet scale $p_0$ and medium independent
cutoff for rescattering) and HIJING~\cite{hijing}.

For simplicity, we have not incorporated any coherent field in the present
study. Including a coherent field in the initial 
condition may decrease the equilibration times and increase
the energy density, the number density and the temperature of the plasma.
We attempt to study the production and equilibration of a QGP
at RHIC and LHC with both coherent field and incoherent partons taken
into account in a forthcoming paper.

\acknowledgements
We thank K.~Kajantie for pointing out an error in the earlier version
of the paper.
This manuscript has been authored under contract No.\ DE-AC02-98CH10886
with the U.S. Department of Energy.
G.C.N.\ acknowledges support by the Alexander von Humboldt Foundation.
A.D.\ is supported from the DOE Research Grant Contract
No.\ De-FG-02-93ER-40764.
%

\newpage

\subsection*{Figure captions}

FIG. 1. Time evolution of the energy density of the gluon-minijets
at RHIC. The solid line is obtained from the self-consistent
solution of the transport equation. The dashed line correspond to the free
streaming and hydrodynamic limits with the same initial minijet conditions
at $\tau_0=1/p_0$.

FIG. 2. Time evolution of the energy density of the gluon-minijets
at LHC. The solid line is obtained from the self-consistent
solution of the transport equation. The dashed line correspond to the free
streaming and hydrodynamic limits with the same initial minijet conditions
at $\tau_0=1/p_0$.

FIG. 3. Time evolution of the temperatures of the gluon-minijets
at RHIC and LHC. 

FIG. 4. Time evolution of the number densities of the gluon-minijets
at RHIC and LHC. 

FIG. 5. Time evolution of the in-medium transport
cross sections for the $gg \rightarrow gg$ process at RHIC and LHC.

FIG. 6. Time evolution of the kinetic relaxation time $\tau_c(\tau)$
at RHIC and LHC.

FIG. 7. Time evolution of the strong coupling constant $\alpha_s(\tau)$ 
at RHIC and LHC.

\newpage

\begin{table}[tb]
\medskip
\begin{tabular}{lcccccc}
& $\sqrt{s}$ (GeV) & $p_0$ (GeV) & A-B & $n_0$ (fm$^{-3}$) 
& $\epsilon_0$ (GeV/fm$^3$) & $T_0$ (GeV) \\ \hline
RHIC & 200 & 1.13 & Au-Au & 34.3 & 56 & 0.535 \\ \hline
LHC & 5500 & 2.13 & Pb-Pb & 321.6 & 1110 & 1.13\\
\end{tabular}
\caption{Initial conditions for the pre-equilibrium evolution of the
plasma, as obtained from the number and energy of gluon-minijets with
$p_t>p_0$.}
\end{table}

\begin{table}[tb]
\medskip
\begin{tabular}{lcccc}
RHIC & $\tau$ (fm) & $\alpha$ (for $\epsilon(\tau)$) & $\alpha$ 
(for T($\tau$)) & $\alpha$ (for n($\tau$)) \\ \hline
 & 0.5  & 1.13 & 0.286 & 0.80   \\ \hline
 & 1.0  & 1.15 & 0.290 & 0.76  \\ \hline
 & 1.5 & 1.17 & 0.295 & 0.75 \\ \hline
 & 2.0 & 1.2 & 0.302 & 0.76  \\ \hline
 & 2.5 & 1.22 & 0.305 & 0.77  \\ \hline
 & 3.0 & 1.23 & 0.31 & 0.782 \\ \hline
 & 3.5 & 1.25 & 0.312 & 0.8  \\ \hline
 & 4.0 & 1.26 & 0.315 & 0.83   \\ \hline 
 & 4.5 & 1.27 & 0.32 & 0.85   \\  
\end{tabular}
\caption{Exponents $\alpha$ obtained by fitting our results with the
functional form $\tau^{-\alpha}$ at RHIC.}
\end{table}

\newpage

\begin{table}[tb]
\medskip
\begin{tabular}{lcccc}
LHC & $\tau$ (fm) & $\alpha$ (for $\epsilon(\tau)$) & $\alpha$ 
(for T($\tau$)) & $\alpha$ (for n($\tau$)) \\ \hline
 & 0.5  & 1.118  & 0.29  & 0.774  \\ \hline
 & 1.0  & 1.164 & 0.297 & 0.75    \\ \hline
 & 1.5 & 1.185 & 0.302 &  0.751  \\ \hline
 & 2.0 & 1.21 & 0.306 &  0.77  \\ \hline
 & 2.5 & 1.227 & 0.308 & 0.8  \\ \hline
 & 3.0 & 1.240 & 0.31 & 0.825   \\ \hline
 & 3.5 & 1.251 & 0.314 & 0.85   \\ \hline
 & 4.0 & 1.262 & 0.317 & 0.875   \\  \hline
 & 4.5 & 1.271 & 0.318 & 0.897  \\  \hline
 & 5.0 & 1.28 & 0.32 & 0.92   \\  \hline
 & 5.5 & 1.28 & 0.32 & 0.92  \\  \hline
 & 6.0 & 1.28 & 0.32 & 0.93  \\  
\end{tabular}
\caption{Exponents $\alpha$ obtained by fitting our results with the
functional form $\tau^{-\alpha}$ at LHC.}
\end{table}

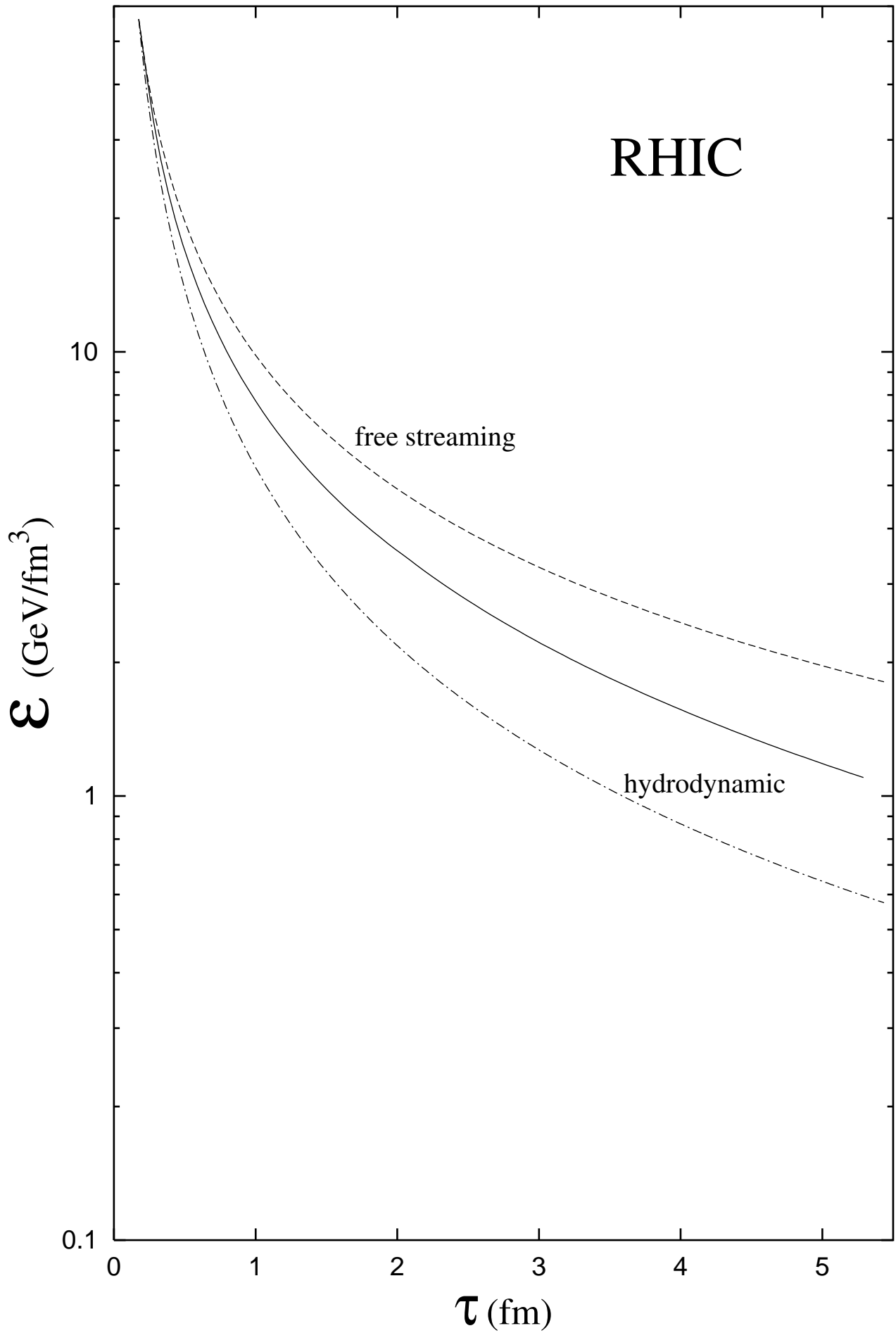
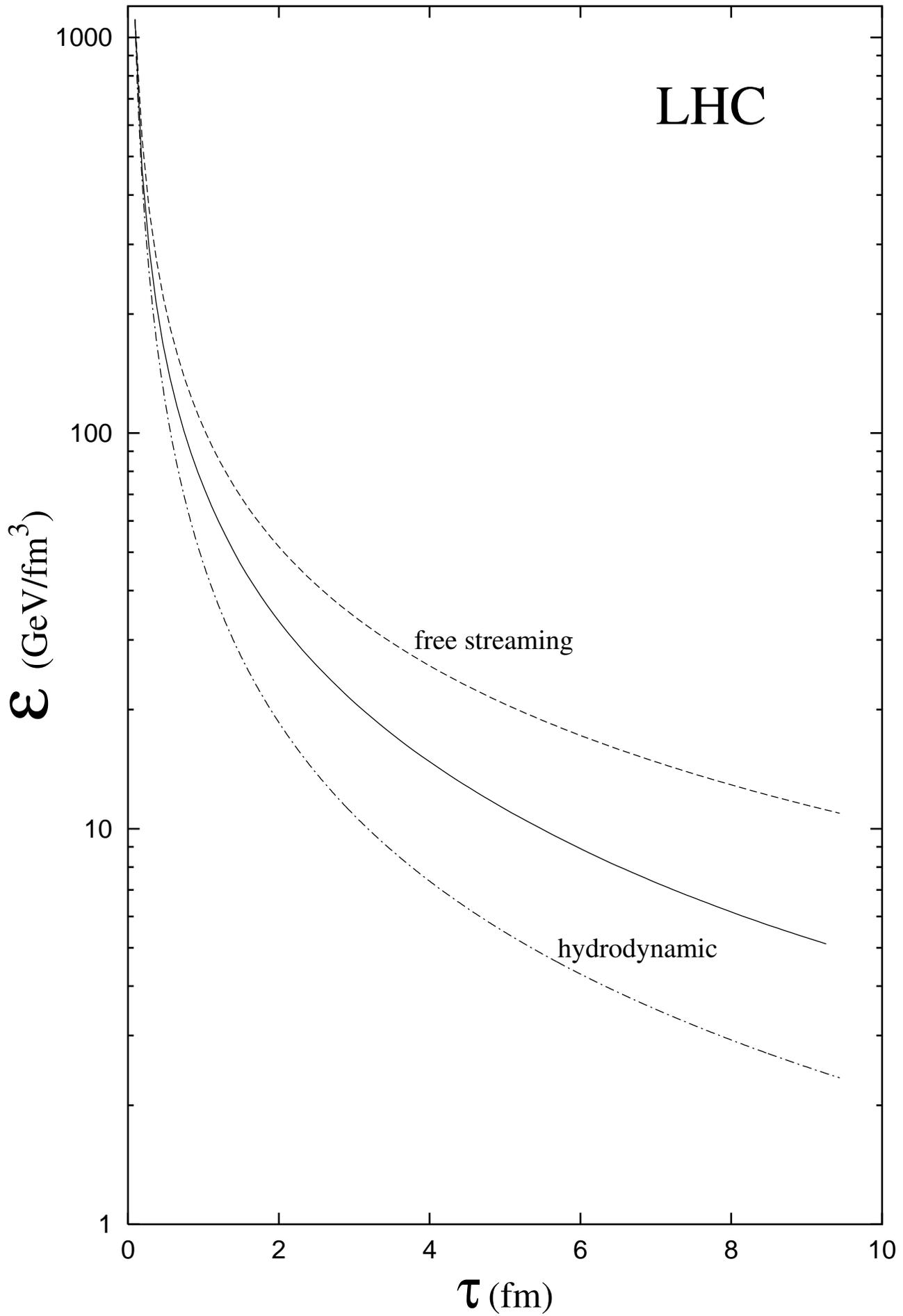
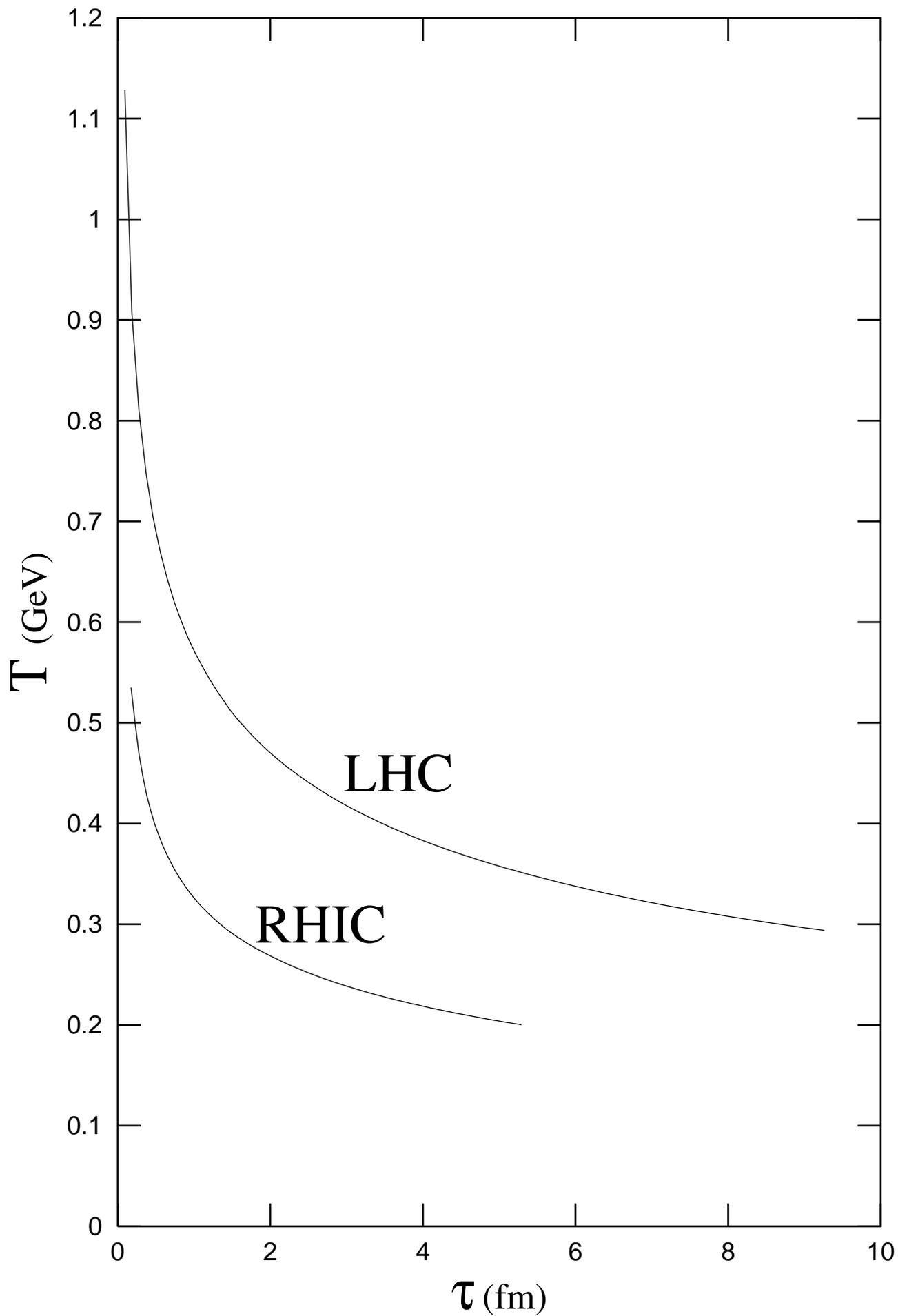
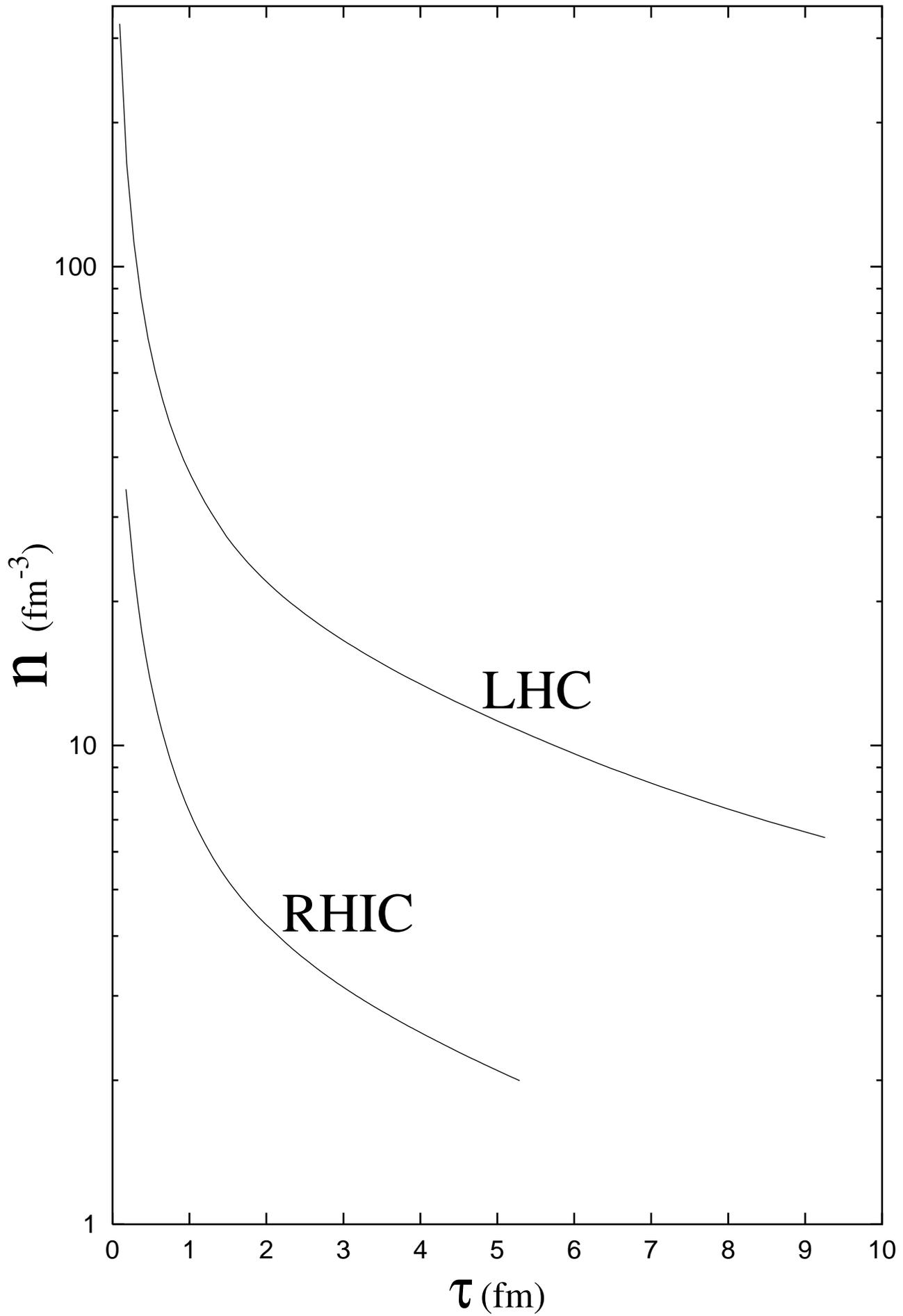
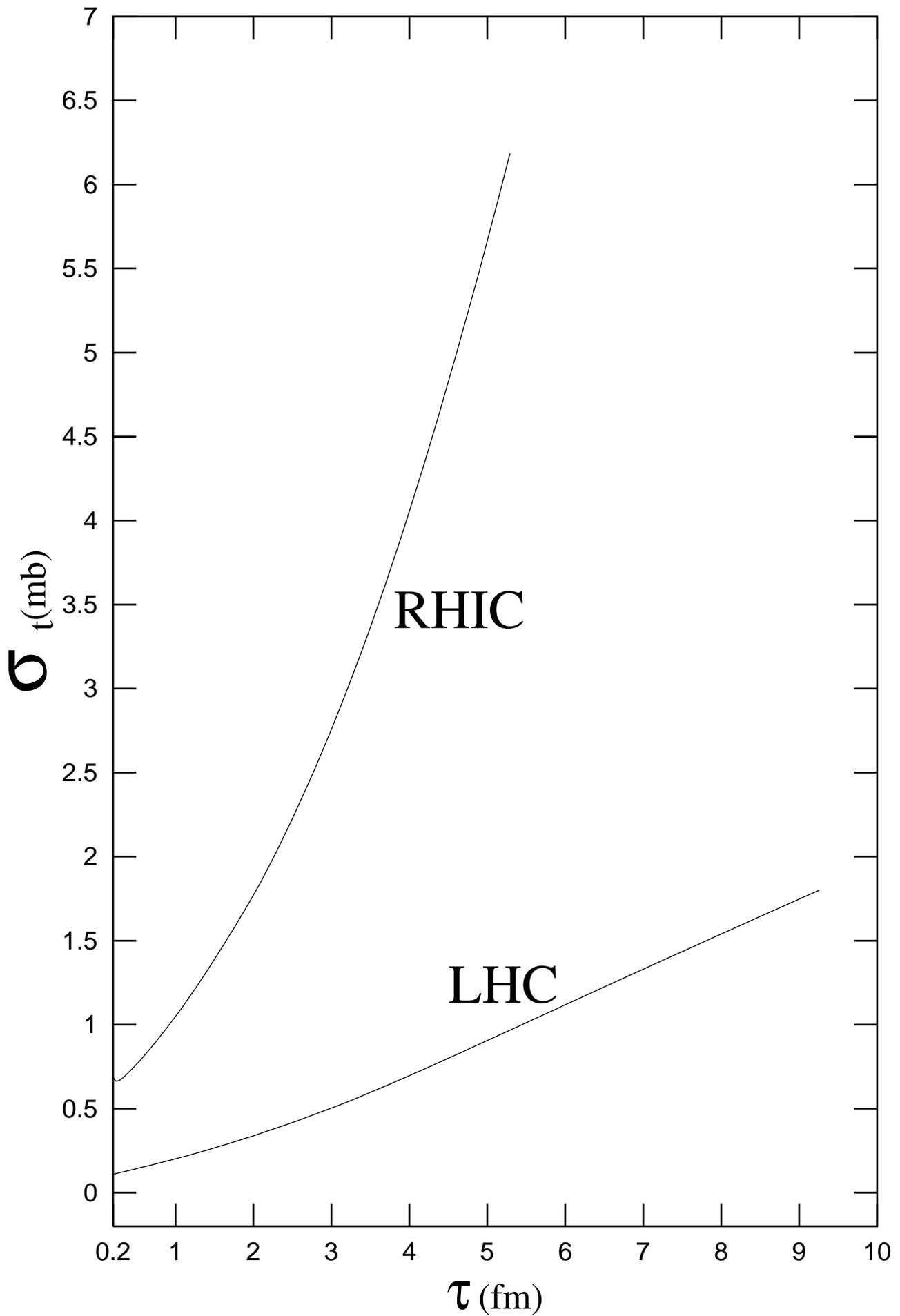
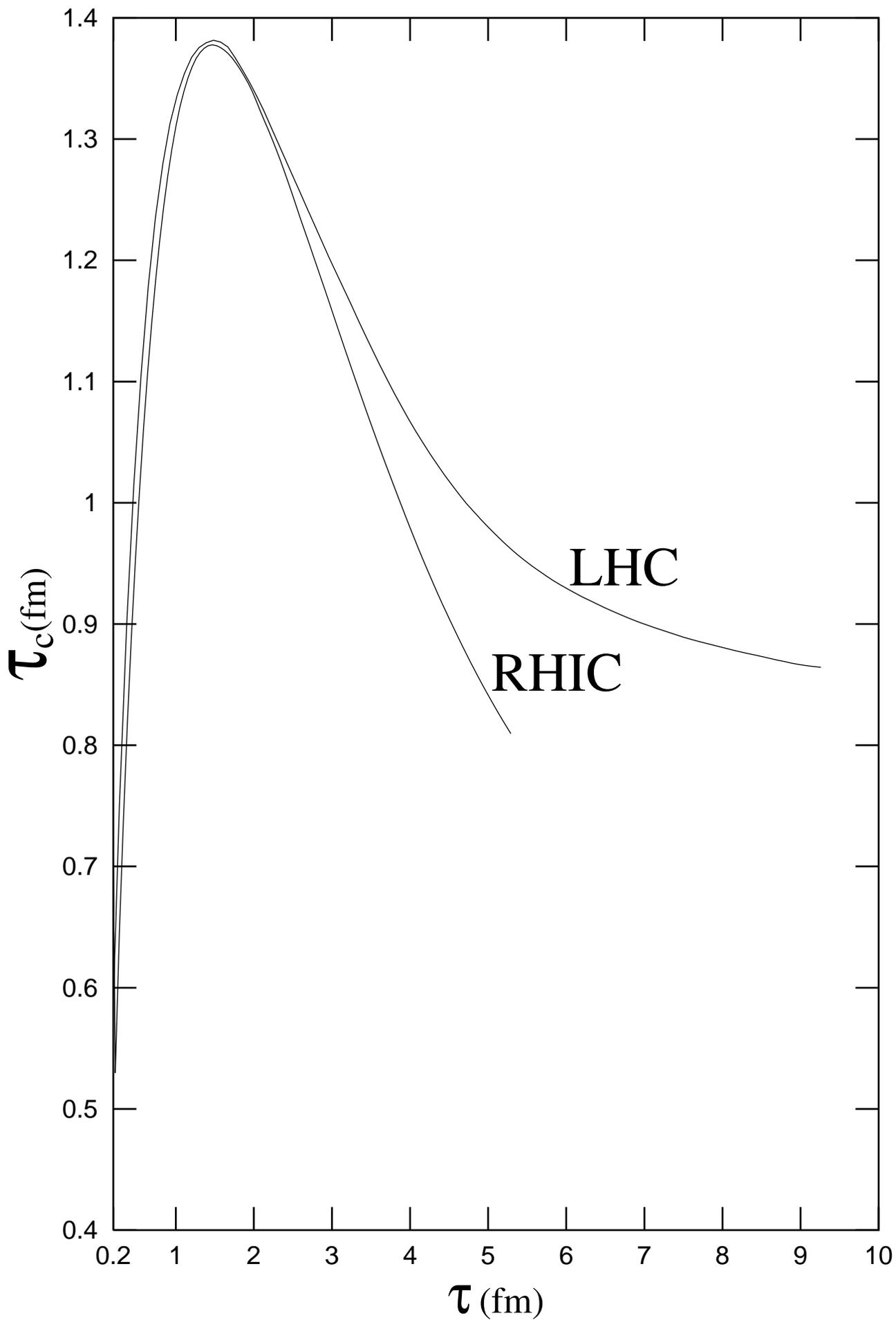
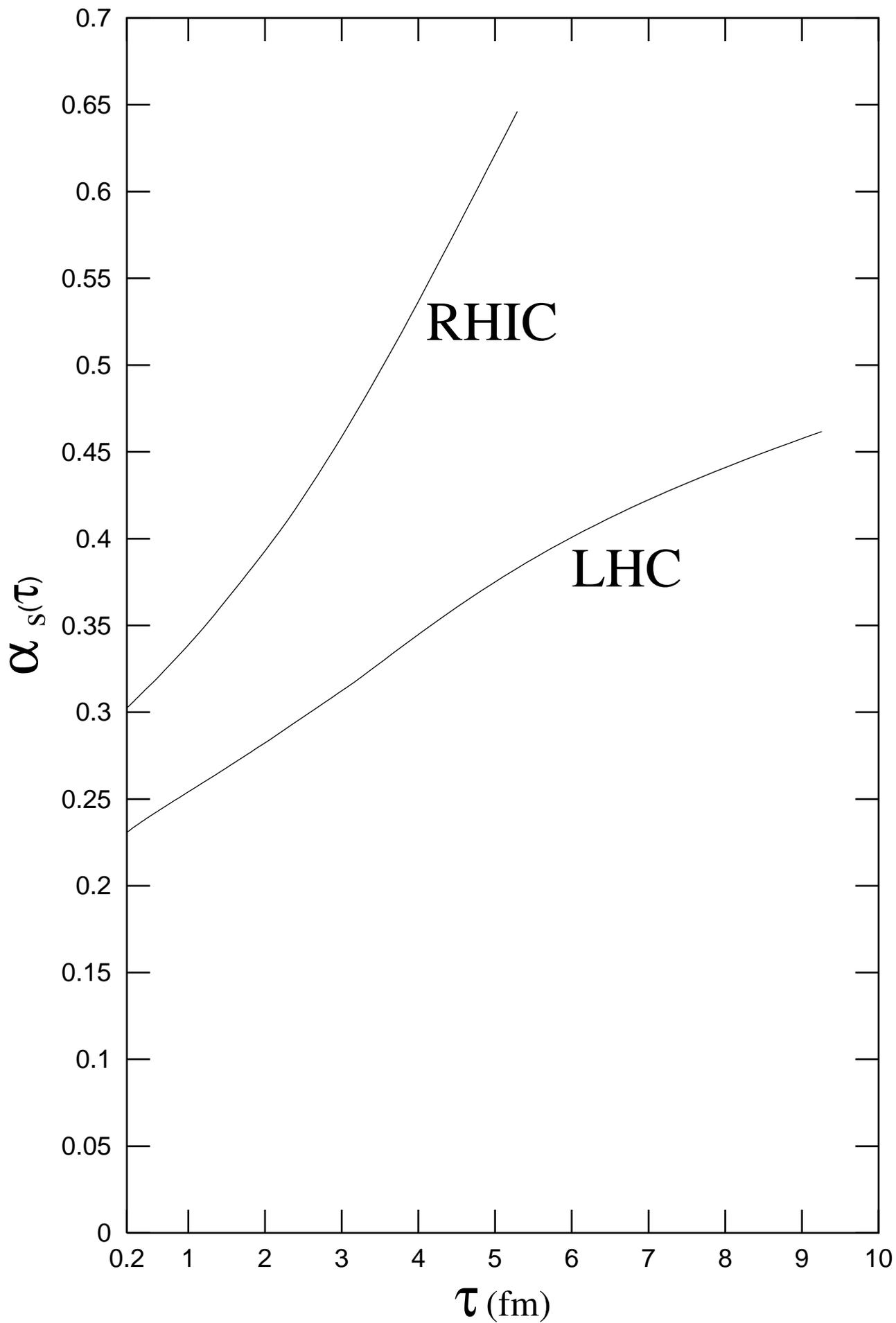
\end{document}